\documentclass[12pt]{article}

\usepackage{amsmath,amssymb}
\usepackage{graphicx}

\newcommand{\ket}[1]{| #1 \rangle}
\newcommand{\bra}[1]{\langle #1 |}
\newcommand{\expv}[1]{\langle #1 \rangle}
\newcommand{\sys}{\mathrm{s}}
\newcommand{\env}{\mathrm{e}}
\newcommand{\prob}{\mathrm{p}}
\newcommand{\uni}{\mathrm{u}}
\newcommand{\Tr}{\mathrm{Tr}}
\newcommand{\kb}{k_\mathrm{B}}

\begin{document}

\begin{center}
{\Large\bf Probabilistic reversing operation
with fidelity and purity gain for macroscopic quantum superposition}
\vskip .6 cm
Hiroaki Terashima$^{1,2}$ and Masahito Ueda$^{1,2}$
\vskip .4 cm
{\it $^1$Department of Physics, Tokyo Institute of Technology,\\
Tokyo 152-8551, Japan} \\
{\it $^2$CREST, Japan Science and Technology Corporation (JST),\\
Saitama 332-0012, Japan}
\vskip .6 cm
\end{center}

\begin{abstract}
It is shown that a large class of weak disturbances on
macroscopic quantum superpositions can be canceled by
a probabilistic reversing operation on the system.
We illustrate this for spin systems undergoing
an Ising-type interaction with the environment
and demonstrate that both the fidelity to the original
state and the purity of the amended state
can simultaneously be increased by the reversing operation.
A possible experimental scheme to implement our
scheme is discussed.
\end{abstract}

\begin{flushleft}
{\footnotesize
}
\end{flushleft}

\section{Introduction}
The macroscopic quantum superpositions
referred to as Schr\"odinger cat states are of great interest
in the context of the foundations of quantum mechanics~\cite{Legget80}
and are an important resource for
quantum  measurement techniques~\cite{GiLlMa04}.
However, cat states are notoriously vulnerable
to perturbations caused by interactions with the environment.
Even when the interactions are very weak,
cat states can easily be destroyed
due to their highly entangled nature.
The state change can be measured by the fidelity
between the original state and the resulting state,
or by the purity of the resulting state.
To recover the original cat state, we could
employ a quantum error-correcting scheme~\cite{NieChu00},
which, however, requires
a macroscopic number of redundant qubits.

In this paper, we propose a probabilistic scheme
to recover a Schr\"odinger cat state,
based on the concept of physical reversibility
in quantum measurement~\cite{UeImNa96,Ueda97}.
A quantum measurement is said to be
physically reversible if
there exists a reversing operation that
recovers the premeasurement state from 
the postmeasurement state by means of a physical process
with a nonzero probability of success.
Since an interaction with the environment can
be modeled as a kind of quantum measurement,
we may expect that the reversing operation could
recover the original cat state in a probabilistic way.
However, the interaction with the environment differs
from a quantum measurement in that
it does not refer to the outcome of the ``measurement'',
and this ignorance usually leads
to decoherence of the system.
Can the physical reversibility be employed
to recover the original cat state even in this situation?
The previous papers~\cite{UeImNa96,Ueda97,%
UedKit92,Imamog93,Royer94,KoaUed99,TerUed03,TerUed05}
on physical reversibility have not addressed this question.
We here give an affirmative answer.
That is, our scheme provides
a considerable increase in the fidelity
together with an increase in the purity
when the interaction of the system with the environment is weak.
The reversing operation invokes a quantum measurement with postselection
and thus has a probabilistic nature.
However, the probability of success is high and
the reversing operation increases
both the mean fidelity and the mean purity.
We shall describe an explicit model
using spin systems with an Ising-type interaction,
and discuss a possible experimental situation.

This paper is organized as follows.
Section~\ref{sec:scheme} develops a general theory of
reversing operation.
Section~\ref{sec:example} considers an example
of a Schr\"odinger cat state.
Section~\ref{sec:conclude} summarizes our results.

\section{\label{sec:scheme}General Theory of Reversing Operation}
Consider that a system with an initial pure density operator
$\hat{\rho}_\sys=\ket{\psi}\bra{\psi}$,
where the state $\ket{\psi}$ is assumed to be unknown to us,
evolves in time according to the Hamiltonian $\hat{H}_\sys$,
and that the system
is disturbed due to its interaction with the environment. 
Let the initial density operator of the environment,
the Hamiltonian of the environment,
and the interaction Hamiltonian be
$\hat{\rho}_\env$, $\hat{H}_\env$, and $\hat{H}_{\sys;\env}$, respectively.
Then, the state of the system after time $t_\env$ is described
by a reduced density operator
\begin{equation}
  \hat{\rho}_\sys'=\Tr_\env\left[e^{-i\hat{H}t_\env}
  \left(\hat{\rho}_\sys\otimes\hat{\rho}_\env\right)
   e^{i\hat{H}t_\env}\right],
\end{equation}
where $\hat{H}\equiv\hat{H}_\sys+\hat{H}_\env+\hat{H}_{\sys;\env}$
denotes the total Hamiltonian
and $\Tr_\env$ denotes the partial trace over the environment.
According to the quantum operation formalism~\cite{NieChu00},
there exists a set of linear operators $\{\hat{E}_k\}$ such that
$\hat{\rho}_\sys'$ is related to $\hat{\rho}_\sys$ by
\begin{equation}
\hat{\rho}_\sys'=\sum_k \hat{E}_k \,\hat{\rho}_\sys \hat{E}_k^\dagger,
\label{eq:noise}
\end{equation}
where $\hat{E}_k$'s satisfy
\begin{equation}
\sum_k \hat{E}_k^\dagger\hat{E}_k=\hat{I}
\label{eq:conde}
\end{equation}
with $\hat{I}$ being the identity operator.
Since $\hat{E}_k$ is a linear operator,
it can uniquely be decomposed by the polar decomposition into
\begin{equation}
  \hat{E}_k=\hat{\mathcal{U}}_k
        \sqrt{\hat{E}_k^\dagger\hat{E}_k},
\end{equation}
where $\hat{\mathcal{U}}_k$ is a unitary operator.
When the interaction with the environment is weak,
$\sqrt{\hat{E}_k^\dagger\hat{E}_k}$ should be expanded as
\begin{equation}
\sqrt{\hat{E}_k^\dagger\hat{E}_k}\simeq
a_k\left(\hat{I}+g\hat{\epsilon}_k^{(1)}
  +g^2\hat{\epsilon}_k^{(2)}\right),
\end{equation}
where $a_k$ is a positive number,
$g$ is a real dimensionless small parameter characterizing
the strength of the interaction between the system and the environment,
and $\hat{\epsilon}_k^{(1)}$ and $\hat{\epsilon}_k^{(2)}$ are
Hermitian operators.
On the other hand, the unitary operator
$\hat{\mathcal{U}}_k$ can, in general,  be written as
\begin{equation}
  \hat{\mathcal{U}}_k\simeq e^{i\gamma_k}e^{ig\hat{\Gamma}_k}
  \equiv e^{i\gamma_k}\,\hat{U}_k,
\end{equation}
where $\gamma_k$ is a real number and
$\hat{\Gamma}_k$ is a Hermitian operator.
Note that $g\hat{\Gamma}_k$ is not necessarily small
even if the interaction is weak
due to the large degrees of freedom of the system and the environment.
The weak interaction only implies that
$\hat{U}_k$ does not depend strongly on $k$.
Then, $\hat{U}_k$ can be decomposed into
\begin{equation}
 \hat{U}_k=e^{ig\hat{\Gamma}_k} \simeq
   e^{ig\hat{\Gamma}}e^{ig^2\hat{\delta}_k^{(2)}},
\label{eq:gamma}
\end{equation}
where $g\hat{\Gamma}$ is a large Hermitian operator
with no dependence on $k$,
and $g^2\hat{\delta}_k^{(2)}$ is a small Hermitian operator.
Thus for the case of a weak interaction, we obtain
\begin{equation}
  \hat{E}_k\simeq a_k \,e^{i\gamma_k} \,\hat{U}_k
   \left(\hat{I}+g\hat{\epsilon}_k^{(1)}
  +g^2\hat{\epsilon}_k^{(2)}\right)
\label{eq:forme}
\end{equation}
with
\begin{equation}
  \hat{U}_k^\dagger\hat{U}_{k'}\simeq \hat{I}
  -ig^2\left(\hat{\delta}_k^{(2)}-\hat{\delta}_{k'}^{(2)}\right).
\label{eq:weakint}
\end{equation}
It follows from Eq.~(\ref{eq:conde}) that
\begin{align}
 & \sum_k a_k^2=1, \label{eq:condep0} \\
 & \sum_k a_k^2\,\hat{\epsilon}_k^{(1)} =0,\label{eq:condep1} \\
 & \sum_k a_k^2\left[(\hat{\epsilon}_k^{(1)})^2
+2\hat{\epsilon}_k^{(2)}\right]=0. \label{eq:condep2}
\end{align}

The crucial observation here is that
for the weak interaction, $\hat{E}_k$ has
an approximate bounded left inverse
and therefore fulfills the condition of
physical reversibility~\cite{UeImNa96,Ueda97}.
In fact, to an accuracy of order $g$, the left inverse of $\hat{E}_k$
can be written as
\begin{equation}
 \hat{E}_k^{-1}\simeq a_k^{-1} \,e^{-i\gamma_k} \,
   \left(\hat{I}-g\hat{\epsilon}_k^{(1)}\right)\, \hat{U}_k^\dagger.
\end{equation}
It should be emphasized that
such a weak interaction with the environment
can profoundly disturb a Schr\"odinger cat state
due to the effect of $\hat{U}_k$.
The extent to which the state of the system is disturbed
can be evaluated quantitatively in terms of the fidelity
of $\hat{\rho}_\sys'$
to $\hat{\rho}_\sys=\ket{\psi}\bra{\psi}$,
$F'\equiv\sqrt{\expv{\hat{\rho}_\sys'}}$,
where $\expv{\hat{O}}\equiv\bra{\psi}\hat{O}\ket{\psi}$.
It is estimated to be
\begin{equation}
F'^2\simeq \sum_k a_k^2\, |\expv{\hat{U}_k}|^2,
\end{equation}
where the small terms
$g\hat{\epsilon}_k^{(1)}$ and $g^2\hat{\epsilon}_k^{(2)}$
are ignored.
Note that we cannot expand $\hat{U}_k$ in terms of $g$,
since $g\hat{\Gamma}_k$ in Eq.~(\ref{eq:gamma})
is, in general, not small.
This means that
the value of $F'$ can be almost $0$ for a cat state
even if $g$ is small.
The state of the system can thus be altered drastically
by the interaction with the environment, however weak it is.
On the other hand,
the extent to which the state of the system
becomes mixed can be quantified by the purity of the system,
$P'\equiv\Tr\hat{\rho}_\sys'^2$,
which is estimated to be
\begin{equation}
P'\simeq 1-2g^2 \sum_k a_k^2\,
\expv{\,(\Delta\hat{\epsilon}_k^{(1)})^2\,}
\label{eq:pu1}
\end{equation}
using Eqs.~(\ref{eq:forme})--(\ref{eq:condep2}),
where $\Delta\hat{O}\equiv \hat{O}-\expv{\hat{O}}$.
The purity thus does not decrease so drastically as the fidelity.

We now discuss the reversing operation to recover
the original state $\hat{\rho}_\sys$ from $\hat{\rho}_\sys'$.
As a simplest example, we consider
the ``average'' of the unitary operators $\{\hat{U}_k^\dagger\}$,
\begin{equation}
\hat{\bar{U}}^\dagger=\sum_k a_k^2\,\hat{U}_k^\dagger,
\label{eq:ubar}
\end{equation}
since each outcome $k$ occurs
with probability $p_k\simeq a_k^2$
by the ``measurement'' done by the environment.
We note that this operator is approximately unitary
for the case of weak interactions.
In fact, given Eq.~(\ref{eq:weakint}), we can easily show that
\begin{equation}
 \hat{\bar{U}}^\dagger\hat{\bar{U}}\simeq \hat{I}.
\end{equation}
By applying the average unitary operator (\ref{eq:ubar})
to the system,
$\hat{\rho}_\sys'$ is changed into
\begin{align}
\hat{\rho}_{\sys,\uni}'' &\equiv
\hat{\bar{U}}^\dagger\hat{\rho}_\sys'\hat{\bar{U}} \notag\\
 &\simeq
 \hat{\rho}_\sys+g^2\sum_k a_k^2\,
 (\hat{\epsilon}_k^{(2)}\hat{\rho}_\sys
  +\hat{\rho}_\sys\hat{\epsilon}_k^{(2)}
 +\hat{\epsilon}_k^{(1)}\hat{\rho}_\sys\,
   \hat{\epsilon}_k^{(1)})
\end{align}
where Eqs.~(\ref{eq:weakint})--(\ref{eq:condep1}) are used.
Thus, up to the first order in $g$,
the original state is recovered
by the reversing operation (\ref{eq:ubar}).
Accordingly, the reversing operation can
increase the fidelity to
\begin{equation}
F_{\uni}''^2\equiv \expv{\hat{\rho}_{\sys,\uni}''}
  \simeq 1-g^2 \sum_k a_k^2\,
\expv{\,(\Delta\hat{\epsilon}_k^{(1)})^2\,}.
\label{eq:funi}
\end{equation}
However, it cannot increase the purity,
\begin{equation}
P_{\uni}''  \equiv \Tr\hat{\rho}_{\sys,\uni}''^2=P',
\end{equation}
since a unitary operation cannot change
the purity of a system.
Therefore, to increase the purity as well as the fidelity,
we must exploit the nonunitary aspect
of the reversing operation.
However, since we assume that we are
ignorant about the state $\ket{\psi}$,
we cannot use the projector $\ket{\psi}\bra{\psi}$
to project the system to $\ket{\psi}$.
Even if we could, it would generally be difficult
to experimentally realize the projection onto the cat state
due to its highly entangled nature.
We will thus need a method for determining
the nonunitary state reduction
without the knowledge of $\ket{\psi}$.

To implement this, consider a situation in which
a probe with initial density operator $\hat{\rho}_\prob$ and
Hamiltonian $\hat{H}_\prob$ interacts
with the system during time $t_\prob$
via an interaction Hamiltonian $\hat{H}_{\sys;\prob}$,
and then the probe is measured with respect to a complete set
of projectors $\{\hat{P}_m\}$
associated with a certain observable.
When the outcome of the probe measurement is $m$,
the postmeasurement state of the system is given by
\begin{equation}
  \hat{\rho}_{\sys|m}''\propto \Tr_\prob
  \left[e^{-i\hat{H}'t_\prob}\left(\hat{\rho}_\sys'\otimes
   \hat{\rho}_\prob\right) e^{i\hat{H}'t_\prob}\hat{P}_m\right],
\label{eq:mea}
\end{equation}
where $\hat{H}'$ is the total Hamiltonian
$\hat{H}'=\hat{H}_\sys+\hat{H}_\prob+\hat{H}_{\sys;\prob}$
and $\Tr_\prob$ denotes the partial trace over the probe.
According to a general formalism of quantum measurement~\cite{NieChu00},
Eq.~(\ref{eq:mea}) can be rewritten
with a set of linear operators $\{\hat{M}_m\}$
called measurement operators as
\begin{equation}
\hat{\rho}_{\sys|m}''=\frac{1}{p_m} \hat{M}_m \,\hat{\rho}_\sys' 
\hat{M}_m^\dagger
\label{eq:measurement}
\end{equation}
with
\begin{equation}
\sum_m \hat{M}_m^\dagger\hat{M}_m=\hat{I},
\label{eq:condm}
\end{equation}
where
\begin{equation}
p_m\equiv\Tr[\,\hat{M}_m \,\hat{\rho}_\sys' 
\hat{M}_m^\dagger\,]
\label{eq:probam}
\end{equation}
is the probability for outcome $m$.
Conversely, we can always construct a probe
that accomplishes a quantum measurement described by
a given set of operators $\{\hat{M}_m\}$
by appropriately choosing $\hat{\rho}_\prob$, $\hat{H}_\prob$,
$\hat{H}_{\sys;\prob}$, and $\{\hat{P}_m\}$.
We here choose them so that for a particular outcome $m_0$,
the corresponding measurement operator is given by
\begin{equation}
  \hat{M}_{m_0}\simeq b
   \left(\hat{I}+\zeta\, g^2\hat{\bar{\epsilon}}^{(2)}\right)
   \hat{\bar{U}}^\dagger,
\label{eq:formm}
\end{equation}
where $b$ is a complex number,
$\zeta$ is a positive number, and
\begin{equation}
 \hat{\bar{\epsilon}}^{(2)}\equiv
   \sum_k a_k^2\, \hat{\epsilon}_k^{(2)}
\label{eq:epbar}
\end{equation}
is the average of $\{\hat{\epsilon}_k^{(2)}\}$.
The other $\hat{M}_m$'s are arbitrary
as long as Eq.~(\ref{eq:condm}) is satisfied.
The measurement operator (\ref{eq:formm}) is
the average unitary operator (\ref{eq:ubar})
followed by the average of
the nonunitary part in Eq.~(\ref{eq:forme}),
except for inclusion of a numerical factor of $\zeta$.
Note that the average of $\{\hat{\epsilon}_k^{(1)}\}$
vanishes because of Eq.~(\ref{eq:condep1}).
The added nonunitary operator
causes state reduction in almost the same way as
the nonunitary operator in Eq.~(\ref{eq:forme}),
purifying the state of the system.
Substituting Eq.~(\ref{eq:formm}) into Eq.~(\ref{eq:probam}),
we obtain
\begin{equation}
 p_{m_0} \simeq |b|^2 \bigl[\,1+2\zeta g^2
 \expv{\hat{\bar{\epsilon}}^{(2)}}\,\bigr].
\label{eq:probm0}
\end{equation}
The positive number $\zeta$ enhances
the probability of obtaining the outcome $m_0$ and
hence the effect of purification.
Substituting Eqs.~(\ref{eq:formm}) and (\ref{eq:probm0})
into Eq.~(\ref{eq:measurement}),
we find
\begin{align}
 \hat{\rho}_{\sys|m_0}'' &\simeq
   \hat{\rho}_\sys+g^2\sum_k a_k^2\,
 (\hat{\epsilon}_k^{(2)}\hat{\rho}_\sys
  +\hat{\rho}_\sys\hat{\epsilon}_k^{(2)}
 +\hat{\epsilon}_k^{(1)}\hat{\rho}_\sys\,
   \hat{\epsilon}_k^{(1)}) \notag \\
 &\qquad+\zeta g^2(\Delta\hat{\bar{\epsilon}}^{(2)}\hat{\rho}_\sys
  +\hat{\rho}_\sys\,\Delta\hat{\bar{\epsilon}}^{(2)}).
\end{align}
Note that $\Tr\hat{\rho}_{\sys|m_0}''=\Tr\hat{\rho}_\sys=1$ holds
because of Eq.~(\ref{eq:condep2}).
The reversing operation (\ref{eq:formm})
also cancels the disturbance up to
the order of $g$.
Calculated up to the order of $g^4$
(see Appendix \ref{sec:fourth}),
the fidelity
$F''_{m_0}\equiv\sqrt{\expv{\hat{\rho}_{\sys|m_0}''}}$
becomes
\begin{align}
F''^2_{m_0} -F_{\uni}''^2 &\simeq 
-\zeta^2 g^4 \expv{\,(\Delta\hat{\bar{\epsilon}}^{(2)})^2\,}
-\frac{1}{2}\zeta g^4\sum_k a_k^2\,\Bigl[\,4\expv{\Delta\hat{\epsilon}_k^{(1)}
\Delta\hat{\bar{\epsilon}}^{(2)}\Delta\hat{\epsilon}_k^{(1)}}  \notag \\
&\qquad\qquad -\expv{\Delta\hat{\bar{\epsilon}}^{(2)}
(\Delta\hat{\epsilon}_k^{(1)})^2}
-\expv{(\Delta\hat{\epsilon}_k^{(1)})^2
 \Delta\hat{\bar{\epsilon}}^{(2)}}\, \Bigr].
\label{eq:difff}
\end{align}
Up to the order of $g^2$, $F_{m_0}''$ is equal to $F_{\uni}''$
given by Eq.~(\ref{eq:funi}).
Thus, if we obtain the measurement outcome $m_0$,
we can increase the fidelity
as in the reversing operation (\ref{eq:ubar}).
Moreover, we can increase the purity as well
in the case of the reversing operation (\ref{eq:formm}).
The purity
$P''_{m_0}\equiv\Tr\hat{\rho}_{\sys|m_0}''^2$ satisfies
\begin{align}
P''_{m_0}-P_{\uni}''= P''_{m_0}-P' &\simeq 
-4\zeta g^4\sum_k a_k^2\,\expv{\Delta\hat{\epsilon}_k^{(1)}
\Delta\hat{\bar{\epsilon}}^{(2)}\Delta\hat{\epsilon}_k^{(1)}} \notag \\
&=2\zeta g^4\sum_{k,k'} a_k^2 a_{k'}^2\,\expv{\Delta\hat{\epsilon}_k^{(1)}
\Delta(\hat{\epsilon}_{k'}^{(1)})^2\Delta\hat{\epsilon}_k^{(1)}}
\label{eq:diffp}
\end{align}
using Eq.~(\ref{eq:condep2}).
Although the increase in the purity is very small
compared to $1$,
its ratio to the lost purity by the environment is
of the order of $g^2$, since
\begin{equation}
 R_\mathrm{p} \equiv \frac{P_{m_0}''-P'}{1-P'}
  \simeq \zeta\times O(g^2)
\end{equation}
from Eq.~(\ref{eq:pu1}).
It is this ratio that is relevant to
an increase in decoherence time.
That is, $R_\mathrm{p}$ increases the
decoherence time from $\tau_\mathrm{d}$
to $\tau_\mathrm{d}/(1-R_\mathrm{p})$.
On the other hand,
the fidelity is increased when $F'\simeq0$ to
\begin{equation}
  R_\mathrm{f}\equiv \frac{F_{m_0}''-F'}{1-F'}
  \simeq 1-O(g^2),
\end{equation}
which is independent of $\zeta$.

\section{\label{sec:example}System of Spin-$1/2$ Particles}

\subsection{Description of the model}
As a concrete example, we consider a system of $N_s$ ($\equiv 2s$)
spin-$1/2$ particles (or two-level systems),
and assume that the system is in a cat state
in which the spin states of the particles
are either all up or all down along the $x$-axis,
\begin{equation}
 \ket{\psi}=
  c_+\ket{\uparrow_x,\uparrow_x,\ldots,\uparrow_x}
  +c_-\ket{\downarrow_x,\downarrow_x,\ldots,\downarrow_x},
\end{equation}
where $|c_+|^2+|c_-|^2=1$.
It is assumed that
we have no \textit{a priori} information about $c_\pm$.
With the spin operators of the $a$-th particle
$\{\hat{s}_{x}^{(a)},\hat{s}_{y}^{(a)},\hat{s}_{z}^{(a)}\}$,
the total spin operator of the system is given by
$\hat{S}_i=\sum_{a=1}^{N_s} \hat{s}_i^{(a)}$,
where $i=x,y,z$.
Let $\ket{s,\sigma}$ be the simultaneous eigenstate of
$\sum_i \hat{S}^2_i$ and $\hat{S}_z$
with eigenvalues $s(s+1)$ and $\sigma$, respectively.
The density operator of the system is then written as
\begin{equation}
\hat{\rho}_\sys=\ket{\psi}\bra{\psi}
  =\sum_{\sigma,\sigma'} c_\sigma c_{\sigma'}^\ast\,
  \ket{s,\sigma}\bra{s,\sigma'},
\end{equation}
where
\begin{equation}
  c_\sigma=\frac{1}{2^{s}}
 \sqrt{\frac{(2s)!}{(s+\sigma)!(s-\sigma)!}}\,
 \left[\,c_++(-1)^{s-\sigma}\,c_- \right].
\label{eq:defc}
\end{equation}
In the following discussions,
we assume that $\hat{H}_\sys=0$ for simplicity.
The environment is assumed to
consist of $N_j$ ($\equiv 2j$) spin-$1/2$ particles
(or two-level systems).
The Hamiltonian of the environment is assumed to be
\begin{equation}
 \hat{H}_\env =-\Delta E\sum_{n=1}^{N_j} \hat{j}_z^{(n)},
 \label{eq:henv}
\end{equation}
where $\Delta E$ is the energy difference
between the spin states
$\ket{\uparrow\,}_n$ and $\ket{\downarrow\,}_n$,
and $\hat{j}_{z}^{(n)}$ is the spin $z$-component operator of
the $n$-th particle of the environment.
The spin of each particle is up with probability $\cos^2(\theta/2)$
and down with probability $\sin^2(\theta/2)$ along the $z$-axis,
where $0\le\theta\le\pi$.
The environment is thus described by the density operator,
\begin{equation}
\hat{\rho}_\env = \prod_{n=1}^{N_j}\left(\,
   \cos^2\frac{\theta}{2}\,\ket{\uparrow\,}_n{}_n\bra{\,\uparrow}
  +\sin^2\frac{\theta}{2}\,\ket{\downarrow\,}_n{}_n\bra{\,\downarrow}
   \,\right).
\label{eq:rhoe}
\end{equation}
The interaction between the system and the environment
is assumed to be
\begin{equation}
 \hat{H}_{\sys;\env} =\alpha_\env
\sum_{n=1}^{N_j} \hat{j}_z^{(n)}\hat{S}_{z},
\end{equation}
where $\alpha_\env$ is a real constant.

\begin{figure}
\begin{center}
\includegraphics[scale=0.65]{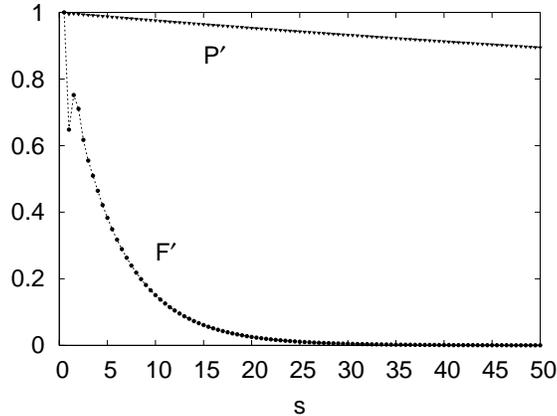}
\end{center}
\caption{\label{fig1}Fidelity $F'$ and purity $P'$
after an interaction with the environment
as functions of spin $s$, which is equal to
one half of the degrees of the system.
The parameters used are
$c_{\pm}=1/\sqrt{2}$, $j=50$, $g=0.01$, and $\theta=\pi/6$.
The fidelity $F'$ oscillates near $s\simeq 1$
due to the effect of statistics
(i.e., $s$ is an integer or a half-integer)
which becomes negligible as $s$ becomes large.}
\end{figure}

The interaction with the environment acts
as a random noise disturbance
on the system when the state of the environment is traced over.
After a certain period of time $t_\env$,
we find that the density operator of the system
is changed from $\hat{\rho}_\sys$ to
\begin{equation}
  \hat{\rho}_\sys'=\sum_{\sigma,\sigma'} 
  c_\sigma c_{\sigma'}^\ast\,
 N_{\sigma\sigma'}^{(j)}(\theta)\,
  \ket{s,\sigma}\bra{s,\sigma'},
\end{equation}
where
\begin{equation}
N_{\sigma\sigma'}^{(j)}(\theta)
=\left(e^{-ig(\sigma-\sigma')}\cos^2\frac{\theta}{2}
    +e^{ig(\sigma-\sigma')}\sin^2\frac{\theta}{2}\right)^{2j}
\label{eq:defn}
\end{equation}
with an effective strength of interaction
$g\equiv \alpha_\env t_\env/2$.
If the interaction is sufficiently weak,
Eq.~(\ref{eq:defn}) can be approximated
up to the order of $g^2$ as
\begin{equation}
N_{\sigma\sigma'}^{(j)}(\theta)
\simeq e^{-g^2j\sin^2\theta\,(\sigma-\sigma')^2-
             i2gj\cos\theta\,(\sigma-\sigma')}.
\label{eq:approxn}
\end{equation}
However, because of the large $s$ and $j$,
the weak interaction strength does not imply
that the perturbation on the system is small.
The fidelity of $\hat{\rho}_\sys'$
to $\hat{\rho}_\sys$ is given by
\begin{equation}
F'^2 = \sum_{\sigma,\sigma'}
     |c_\sigma|^2|c_{\sigma'}|^2\,
  \mathrm{Re} \left[ N_{\sigma\sigma'}^{(j)}(\theta) \right].
\label{eq:def1stf}
\end{equation}
Figure \ref{fig1} shows $F'$
as a function of $s$ for $c_{\pm}=1/\sqrt{2}$,
$j=50$, $g=0.01$, and $\theta=\pi/6$.
The fidelity $F'$ decreases
as the degrees of freedom of the system $2s$ become large.
This decrease results from rapid oscillation of
the cosine factor in
\begin{equation}
 \mathrm{Re} \left[ N_{\sigma\sigma'}^{(j)}(\theta) \right]
   \simeq e^{-g^2j\sin^2\theta\,(\sigma-\sigma')^2}
      \cos\left[\,2gj(\sigma-\sigma')\cos\theta\,\right],
\end{equation}
which represents the change in the relative phase
between $\ket{s,\sigma}$ and $\ket{s,\sigma'}$.
The cosine factor oscillates with period $\pi/gj\cos\theta$
as a function of $\sigma-\sigma'$,
while the weight $|c_\sigma|^2|c_{\sigma'}|^2$
concentrates near $(\sigma,\sigma')=(0,0)$ with width $\sqrt{s}$
owing to the binomial coefficient in Eq.~(\ref{eq:defc}).
Since any oscillating function cancels out
when it is averaged over the argument, we estimate that
\begin{equation}
F'\simeq 0
\end{equation}
if $s$ is so large that
\begin{equation}
s\gtrsim\frac{\pi^2}{4g^2j^2\cos^2\theta}.
\label{eq:fdecay}
\end{equation}
Actually, as long as $s$ is large,
the central limit theorem gives
an analytic expression for the fidelity
as (see Appendix \ref{sec:degf})
\begin{equation}
 F'^2 \sim
\frac{1}{\sqrt{1+2g^2sj\sin^2\theta}}
\exp\left[\frac{-2g^2sj^2\cos^2\theta}{1+2g^2sj\sin^2\theta}\right].
\label{eq:fexpress}
\end{equation}

The interaction with the environment also
causes the purity to degrade.
The purity of $\hat{\rho}_\sys'$ is given by
\begin{equation}
P' =\sum_{\sigma,\sigma'}
  |c_\sigma|^2|c_{\sigma'}|^2\,
  \left| N_{\sigma\sigma'}^{(j)}(\theta)\right|^2.
\end{equation}
Figure \ref{fig1} shows $P'$
as a function of $s$ for $c_{\pm}=1/\sqrt{2}$,
$j=50$, $g=0.01$, and $\theta=\pi/6$.
The purity $P'$ decreases
as the degrees of freedom of the system $2s$ become large,
but this decrease is not so drastic as in the case of the fidelity,
since $|N_{\sigma\sigma'}^{(j)}(\theta)|^2$
involves no cosine factor;
it is calculated to be
\begin{align}
 P'  &\simeq \sum_{\sigma,\sigma'}
  |c_\sigma|^2|c_{\sigma'}|^2\,
  \left[\,1-2g^2j\sin^2\theta\,(\sigma-\sigma')^2\,\right] \notag \\
 &= 1-2g^2sj\sin^2\theta
\label{eq:1stp}
\end{align}
if $s$ is so small that
\begin{equation}
 s\ll \frac{1}{g^2j\sin^2\theta}.
\label{eq:psmall}
\end{equation}
Therefore, the cat state is drastically changed
together with a slight degradation in the purity
of the quantum state
after the interaction with the environment.
Moreover, we know neither
$\hat{\rho}_\sys$ nor $\hat{\rho}_\sys'$
due to our ignorance of $c_\pm$.
Nevertheless, these facts do not imply that
the original cat state cannot be recovered.

As in Eq.~(\ref{eq:noise}), 
$\hat{\rho}_\sys'$ can be written in terms of $\{\hat{E}_k\}$
of the form of Eq.~(\ref{eq:forme}).
We find that
\begin{align}
 a_k &= \frac{1}{2^j}
      \sqrt{\frac{(2j)!}{(j+k)!(j-k)!}}, \label{eq:ak} \\
 \hat{U}_k &= e^{ig\hat{\Gamma}_k}=
  e^{-i2gj\cos\theta\,\hat{S}_z
  -ig^2k\sin 2\theta\,\hat{S}_z^2}, \label{eq:uk} \\
 \hat{\epsilon}_k^{(1)} &=
  -2k\sin\theta\,\hat{S}_z, \\
 \hat{\epsilon}_k^{(2)} &=
 2(k^2-j)\sin^2\theta\,\hat{S}_z^2, \label{eq:epk} \\
 \hat{\delta}_k^{(2)} &=-k\sin 2\theta\,\hat{S}_z^2
\end{align}
for  $k=j,j-1,\ldots,-j+1,-j$.
The condition (\ref{eq:fdecay}) implies
that $g\hat{\Gamma}_k$ is large for the cat state,
while the condition (\ref{eq:psmall}) implies
that $g\hat{\epsilon}_k^{(1)}$ and
$g^2\hat{\epsilon}_k^{(2)}$ are small because of
$\expv{\hat{S}_z^2}=s/2$ and
$\sum_k a_k^2 \,k^2=j/2$.
The latter condition also implies that
$\hat{U}_k$ does not strongly depend on $k$
as in Eq.~(\ref{eq:weakint}), since
\begin{equation}
\hat{U}_k^\dagger\hat{U}_{k'}\simeq \hat{I}+
  ig^2(k-k')\sin 2\theta\,\hat{S}_z^2.
\end{equation}
Below, we will concentrate on the case in which
$s$ satisfies the following conditions:
\begin{equation}
\frac{\pi^2}{4g^2j^2\cos^2\theta} \lesssim s
\ll \frac{1}{g^2j\sin^2\theta}.
\label{eq:conds}
\end{equation}

\subsection{Reversing operation}
\begin{figure}
\begin{center}
\includegraphics[scale=0.65]{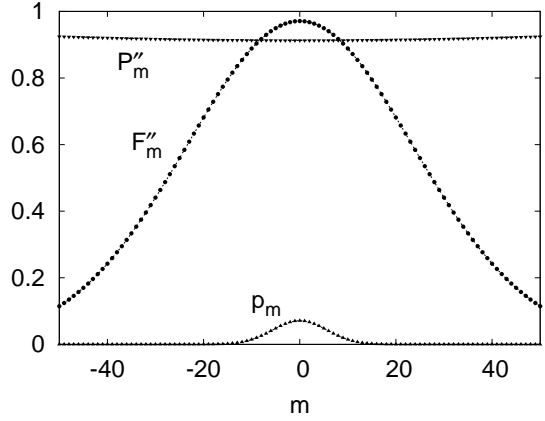}
\end{center}
\caption{\label{fig2}Probability $p_m$, fidelity $F_{m}''$,
and purity $P_{m}''$ after a measurement
as functions of outcome $m$ for
$s=j=50$, $c_{\pm}=1/\sqrt{2}$,
$g=0.01$, and $\theta=\pi/6$.
The mean fidelity and purity after the measurement
are $0.945$ and $0.913$, respectively.}
\end{figure}

To recover the cat state $\hat{\rho}_\sys$,
we perform a measurement~\cite{TerUed05}
on the state $\hat{\rho}_\sys'$
using the information about the environment $(g,j,\theta)$.
The probe of the measurement is a spin-$j$ system
whose spin operators are
$\{\hat{J}_{x},\hat{J}_{y},\hat{J}_{z}\}$
and Hamiltonian is $\hat{H}_\prob=0$.
The measurement proceeds as follows.
The probe is first prepared in a coherent spin state
$\ket{\pi-\theta,\pi/2}$, i.e.,
the eigenstate of the spin component
$\hat{J}_{y}\sin\theta-\hat{J}_{z}\cos\theta$
with eigenvalue $j$~\cite{ACGT72}.
The density operator of the probe is
\begin{equation}
  \hat{\rho}_\prob=\ket{\pi-\theta,\pi/2}\bra{\pi-\theta,\pi/2},
\label{eq:rhop}
\end{equation}
which represents a pure state in contrast with
the state of the environment (\ref{eq:rhoe}).
The probe then interacts with the system
via an interaction Hamiltonian
\begin{equation}
\hat{H}_{\sys;\prob}=\alpha_\prob \hat{J}_z \hat{S}_{z},
\label{eq:hprob}
\end{equation}
where $\alpha_\prob$ is a real constant.
The interaction is turned on during time $t_\prob$
so that $\alpha_\prob t_\prob/2=g$.
After the interaction, a unitary operator
\begin{equation}
\hat{U}_\prob=e^{-i\pi\hat{J}_{y}/2}
\label{eq:uprob}
\end{equation}
is applied to the probe, and finally,
the projective measurement on the probe observable
$\hat{J}_{z}$ is performed.
Let $m$ be the outcome of the measurement,
where $m=j,j-1,\ldots,-j+1,-j$.
The probability for obtaining outcome $m$ is
\begin{equation}
 p_{m} = \sum_{\sigma} |c_\sigma|^2\,
|A_{m\sigma}^{(j)}(\theta)|^2,
\end{equation}
where
\begin{align}
 A_{m\sigma}^{(j)}(\theta)&\equiv \frac{e^{-ij\pi/2}}{2^j}
      \sqrt{\frac{(2j)!}{(j+m)!(j-m)!}}  \notag \\
  &  \qquad{}\times
     \left( e^{-ig\sigma}\sin\frac{\theta}{2}
     +ie^{ig\sigma}\cos\frac{\theta}{2} \right)^{j-m} \notag \\
  &  \qquad{}\times
     \left( e^{-ig\sigma}\sin\frac{\theta}{2}
     -ie^{ig\sigma}\cos\frac{\theta}{2} \right)^{j+m}.
\end{align}
The expectation value of outcome,
$\sum_{m} m p_m$, is calculated to give $0$.
Figure \ref{fig2} shows $p_m$
as a function of $m$ for $s=j=50$, $c_{\pm}=1/\sqrt{2}$,
$g=0.01$, and $\theta=\pi/6$.
The postmeasurement state is given by
\begin{equation}
  \hat{\rho}_{\sys|m}'' =\frac{1}{p_{m}} \sum_{\sigma,\sigma'}
   c_\sigma c_{\sigma'}^\ast\,
   N_{\sigma\sigma'}^{(j)}(\theta)\,
   R_{m\sigma\sigma'}^{(j)}(\theta)\,
  \ket{s,\sigma}\bra{s,\sigma'},
\end{equation}
where
$R_{m\sigma\sigma'}^{(j)}(\theta)=
A_{m\sigma}^{(j)}(\theta)\,A_{m\sigma'}^{(j)\ast}(\theta)$.

As in Eq.~(\ref{eq:measurement}),
the measurement is described by measurement operators
\begin{equation}
\hat{M}_{m}=\sum_\sigma
A_{m\sigma}^{(j)}(\theta)\,\ket{s,\sigma}\bra{s,\sigma}
\end{equation}
with $m=j,j-1,\ldots,-j+1,-j$.
The particular outcome $m_0$
in Sec.~\ref{sec:scheme}
is the expectation value of outcome $m=0$.
In fact, when expanded up to the second order of $g$,
the operator $\hat{M}_{0}$ has the form
of Eq.~(\ref{eq:formm}) with
\begin{align}
b &=\frac{e^{-ij\pi/2}}{2^j}
      \sqrt{\frac{(2j)!}{j!j!}}, \\
\zeta &= 2,
\end{align}
since
\begin{align}
 \hat{\bar{\epsilon}}^{(2)} &= -j\sin^2\theta\,\hat{S}_z^2, \\
 \hat{\bar{U}}^\dagger &= e^{i2gj\cos\theta\,\hat{S}_z},
 \label{eq:ubar2}
\end{align}
from the definitions (\ref{eq:ubar}) and (\ref{eq:epbar})
with Eqs.~(\ref{eq:ak}), (\ref{eq:uk}),
and (\ref{eq:epk}).

The fidelity of $\hat{\rho}_{\sys|m}''$
to $\hat{\rho}_\sys$ is
\begin{align}
F_{m}''^2 &\equiv\expv{\hat{\rho}_{\sys|m}''} \\
  &= \frac{1}{p_{m}} \sum_{\sigma,\sigma'}
    |c_\sigma|^2|c_{\sigma'}|^2\,
  \mathrm{Re} \left[ N_{\sigma\sigma'}^{(j)}(\theta)\,
   R_{m\sigma\sigma'}^{(j)}(\theta) \right],
\end{align}
and, in particular, the fidelity
at the expectation value $m=0$ is
\begin{equation}
 F_{0}''^2  \simeq 1-g^2sj\sin^2\theta
\end{equation}
using the expansions of $p_{m}$,
$N_{\sigma\sigma'}^{(j)}(\theta)$, and
$R_{m\sigma\sigma'}^{(j)}(\theta)$
up to the order of $g^2$.
The mean squared fidelity and the mean fidelity are
given by
\begin{align}
\sum_m p_{m}F_{m}''^2   &\simeq 1-2g^2sj\sin^2\theta, \\
\sum_m p_{m}F_{m}''   &\simeq 1-g^2sj\sin^2\theta.
\end{align}
The fidelity $F_{m}''$ is also shown in Fig.~\ref{fig2}
as a function of $m$ when $s=j=50$, $c_{\pm}=1/\sqrt{2}$,
$g=0.01$, and $\theta=\pi/6$.
In this case, even though
the effect of the environment is to decrease
the fidelity to $F'=2.41\times 10^{-4}$,
the reversing measurement recovers it
to $F_{0}''=0.971$ when the most probable outcome $m=0$
is obtained by the measurement.
This means that $97.1\%$ of the lost fidelity
is recovered by the measurement.
Even if $m\neq0$, the fidelity is still recovered
as long as $m\simeq0$, which occurs with a high probability.
In fact, the mean fidelity after the measurement,
$\sum_m p_m F_{m}''$, is $0.945$.
Under the condition (\ref{eq:conds}),
the fidelity is drastically reduced
by the interaction with the environment
but is almost recovered by the reversing measurement.
Surprisingly, such a drastic change occurs
with a weak interaction.
This is due to
the large degrees of freedom
being involved in the cat state.

On the other hand,
the purity
$P_{m}''  \equiv \Tr\hat{\rho}_{\sys|m}''^2$
is calculated to be
\begin{align}
P_{m}''  &=\frac{1}{p_{m}^2}\sum_{\sigma,\sigma'}
    |c_\sigma|^2|c_{\sigma'}|^2\,
  \left| N_{\sigma\sigma'}^{(j)}(\theta)\,
   R_{m\sigma\sigma'}^{(j)}(\theta)\right|^2 \\
   &\simeq  1-2g^2sj\sin^2\theta,
\end{align}
which is equal to $P'$ in Eq.~(\ref{eq:1stp})
up to the second order of $g$,
independently of $m$.
However, Eq.~(\ref{eq:diffp}) shows that
the purity is indeed increased by the measurement
with outcome $m=0$,
\begin{equation}
 P_{0}''-P'  \simeq 8g^4s^2j^2\sin^4\theta>0.
\label{eq:incp}
\end{equation}
Its ratio to the lost purity by the environment is given by
\begin{equation}
 R_\mathrm{p}= \frac{P_{0}''-P'}{1-P'}
  \simeq 4g^2sj \sin^2\theta.
\end{equation}
The purity $P_{m}''$ is shown in Fig.~\ref{fig2}
as a function of $m$ for $s=j=50$, $c_{\pm}=1/\sqrt{2}$,
$g=0.01$, and $\theta=\pi/6$.
The effect of the environment lowers
the purity to $P'=0.895$,
but the measurement with outcome $m=0$ recovers
it to $P_{0}''=0.913$.
Therefore, $R_\mathrm{p}$ is equal to $0.172$,
indicating that $17.2\%$ of
the lost purity is recovered by the reversing measurement.
The measurement can thus
increase the decoherence time by about $20.7\%$.
The mean purity
\begin{equation}
\sum_m p_m P_{m}'' \simeq 1-2g^2sj\sin^2\theta
\end{equation}
is also increased to $0.913$ by the measurement
as in the case of fidelity.

For comparison, let us examine the degree to which
we can restore the original state
by performing the average unitary operation (\ref{eq:ubar2})
on state $\hat{\rho}_\sys'$.
The resulting state is given by
\begin{equation}
  \hat{\rho}_{\sys,\uni}'' 
 = \sum_{\sigma,\sigma'} c_\sigma c_{\sigma'}^\ast\,
       N_{\sigma\sigma'}^{(j)}(\theta)\,
      e^{i2gj\cos\theta\,(\sigma-\sigma')}\,
      \ket{s,\sigma}\bra{s,\sigma'}.
\end{equation}
The unitary operation
can recover the fidelity like the measurement.
In fact, Eq.~(\ref{eq:difff}) shows that
the fidelity
$F_{\uni}''$ is the same as $F_{0}''$ up to the order of $g^5$,
\begin{equation}
 F_{0}''^2-F_{\uni}''^2 \simeq O(g^6).
\end{equation}
For the previous example
($s=j=50$, $c_{\pm}=1/\sqrt{2}$, $g=0.01$, and $\theta=\pi/6$),
the unitary operation also increases
the fidelity to $F_{\uni}''=0.971$.
Unlike the measurement, however,
the unitary operation cannot increase the purity.
The purity $P_{\uni}''$ is equal to $P'$ and thus satisfies
\begin{equation}
 P_{0}''-P_{\uni}''  \simeq 8g^4s^2j^2\sin^4\theta>0
\end{equation}
as in Eq.~(\ref{eq:incp}).

\subsection{Feasibility of the reversing operation}
To perform the reversing operation, we must know
the parameters of the environment,
such as the effective strength of the interaction $g$,
the number of particles $2j$,
and the probability for spin-up state $\cos^2(\theta/2)$.
From the energy difference $\Delta E$ 
in Eq.~(\ref{eq:henv}),
we can estimate $\theta$
when the environment is in thermal equilibrium
at temperature $T$ by using the relation
\begin{equation}
\tan^2\frac{\theta}{2}=e^{-\Delta E/\kb T},
\end{equation}
where $\kb$ is the Boltzmann constant.
On the other hand,
to estimate $g$ and $j$,
we regard $F'$ and $P'$ as functions of time $t_\env$,
since they depend on $t_\env$
through $g=\alpha_\env t_\env/2$.
Let $t_0$ be the time when $F'$ becomes $0$.
From Eqs.~(\ref{eq:fdecay}) and (\ref{eq:1stp}),
we obtain
\begin{align}
  g^2_0j^2 &\simeq \frac{\pi^2}{4s\cos^2\theta}, \\
  g^2_0j   &\simeq \frac{1-P'_0}{2s\sin^2\theta},
\end{align}
where $g_0$ and $P'_0$ are $g$ and $P'$ at $t_0$,
respectively.
Combining these relations,
we can estimate $j$, $g_0$, and $g=g_0t_\env/t_0$.

Finally, we describe
a possible experimental situation
for the reversing operation~\cite{TerUed05}.
Consider the system as
an ensemble of $2s$ two-level atoms
and the probe as
an ensemble of $2j$ photons with two polarizations
(horizontal or vertical).
The photons can then be regarded as a spin-$j$ system
using the spin operators
\begin{align}
\hat{J}_x &\equiv \frac{1}{2}
   \left(\hat{a}_1^\dagger\hat{a}_2
     +\hat{a}_2^\dagger\hat{a}_1\right), \notag \\
\hat{J}_y &\equiv \frac{1}{2i}
   \left(\hat{a}_1^\dagger\hat{a}_2
     -\hat{a}_2^\dagger\hat{a}_1\right),  \\
\hat{J}_z &\equiv \frac{1}{2}
   \left(\hat{a}_1^\dagger\hat{a}_1
     -\hat{a}_2^\dagger\hat{a}_2\right), \notag
\end{align}
where $\hat{a}_1$ and $\hat{a}_2$ are
the annihilation operator for photons
with horizontal and vertical polarizations, respectively.
The initial probe state (\ref{eq:rhop}) is
prepared by subjecting horizontally polarized photons to
a half-wave plate and a phase shifter, and
the interaction (\ref{eq:hprob}) is realized
by using the paramagnetic Faraday rotation~\cite{KMJYEB99,THTTIY99}.
The unitary operator (\ref{eq:uprob}) corresponds
to a half-wave plate and
the projective measurement of $\hat{J}_{z}$ is achieved
by two photodetectors for the two polarizations.
Typically, in this situation, $s\sim 10^{8}$, $j\sim 10^{8}$,
and $g\sim 10^{-8}$.
Therefore, the condition (\ref{eq:conds})
is satisfied if $\theta^2\ll 1$,
which means $\kb T\ll \Delta E /\ln 4$.

\section{\label{sec:conclude}Conclusion}
We have proposed a probabilistic reversing operation
that can recover both the fidelity and purity after
they are deteriorated through weak interactions with the environment.
Since there is no unitary operation that can increase the purity,
the reversing operation must involve a
nonunitary state reduction of a quantum measurement.
We have considered an ensemble of spin-$1/2$ particles
in a Schr\"odinger cat state as a system and
another ensemble of spin-$1/2$ particles
in a mixed state as an environment.
The cat state of the system is then destroyed by
a weak Ising-type interaction with the environment.
The fidelity to the original cat state
is drastically decreased due to
the large degrees of freedom of the system and environment,
despite a slight decrease in the purity of the system.
We have also shown that a reversing operation
can achieve a profound recovery of the fidelity
together with a nonzero increase in the purity
with a high probability of success.
The reversing operation is achieved by
a quantum measurement that uses a probe.
The probe is a spin system in a coherent spin state
and interacts with the system via an Ising-type Hamiltonian.
If the measurement ends with a preferred outcome,
we can increase not only the fidelity but also the purity
by the reversing operation.

We have also discussed a physical implementation of our model
using two-level atoms as a system and photons as a probe.
Since the interaction would be feasible
in view of recent advances in experimental techniques,
the reversing operation could be experimentally realized
in the near future.
Although in this paper we have focused on
the reversing operation of a Schr\"odinger cat state
undergoing an Ising-type interaction,
our scheme could equally
be applied to other general states undergoing
a large class of weak interactions.

\section*{Acknowledgments}
This research was supported by a Grant-in-Aid
for Scientific Research (Grant No.~17071005) from
the Ministry of Education, Culture, Sports,
Science and Technology of Japan.

\appendix
\section*{Appendix}

\section{\label{sec:fourth}Fourth-Order Calculation}
We here outline the derivation of
Eqs.~(\ref{eq:difff}) and (\ref{eq:diffp}).
To calculate the fidelity and
the purity to the order of $g^4$,
one might think that
Eqs.~(\ref{eq:forme}), (\ref{eq:weakint}), and (\ref{eq:formm})
should be expanded to the order of $g^4$ as
\begin{align}
  \hat{E}_k &\simeq a_k \,e^{i\gamma_k} \,\hat{U}_k
   \left(\hat{I}+g\hat{\epsilon}_k^{(1)}
  +g^2\hat{\epsilon}_k^{(2)}+g^3\hat{\epsilon}_k^{(3)}
   +g^4\hat{\epsilon}_k^{(4)}\right), \\
  \hat{U}_k^\dagger\hat{U}_{k'} &\simeq \hat{I}
  -ig^2\left(\hat{\delta}_k^{(2)}-\hat{\delta}_{k'}^{(2)}\right)
   +g^3\hat{\delta}_{kk'}^{(3)}+g^4\hat{\delta}_{kk'}^{(4)}, \\
  \hat{M}_{m_0} &\simeq b
   \left(\hat{I}+\zeta\, g^2\hat{\bar{\epsilon}}^{(2)}
  +g^3\hat{\chi}^{(3)}+g^4\hat{\chi}^{(4)}\right)
   \hat{\bar{U}}^\dagger,
\end{align}
where $\hat{\epsilon}_k^{(3)}$, $\hat{\epsilon}_k^{(4)}$,
$\hat{\delta}_{kk'}^{(3)}$, $\hat{\delta}_{kk'}^{(4)}$,
$\hat{\chi}^{(3)}$, and $\hat{\chi}^{(4)}$ are
Hermitian operators.
However, we can show that such higher-order terms are irrelevant
to the calculation of $F''^2_{m_0} -F_{\uni}''^2$ and
$P''_{m_0} -P_{\uni}''$ to the order of $g^4$.
The proof goes as follows.
We first consider the contribution of $\hat{\chi}^{(4)}$.
In doing so, we can ignore the other operators,
since $\hat{\chi}^{(4)}$ is of the fourth order itself.
We then obtain
\begin{equation}
\hat{\rho}_{\sys|m_0}''\simeq
 \hat{\rho}_\sys+
 g^4(\Delta\hat{\chi}^{(4)}\hat{\rho}_\sys
 +\hat{\rho}_\sys\,\Delta\hat{\chi}^{(4)}),
\end{equation}
which results in $F''_{m_0}=P''_{m_0}=1$.
Therefore, $\hat{\chi}^{(4)}$ does not contribute
to the fidelity and the purity to the order of $g^4$.
Similarly, $\hat{\chi}^{(3)}$ does not do so to the order of $g^3$.
Although $\hat{\chi}^{(3)}$ may appear in the fourth order
together with $\hat{\epsilon}_k^{(1)}$,
such a term vanishes because of Eq.~(\ref{eq:condep1})
by the summation over $k$.
From these findings, we can set
\begin{equation}
\hat{\chi}^{(3)}=\hat{\chi}^{(4)}=0.
\end{equation}
On the other hand, if $\hat{\bar{\epsilon}}^{(2)}=0$,
$F''^2_{m_0} -F_{\uni}''^2=P''_{m_0} -P_{\uni}''=0$,
since $\hat{M}_{m_0}$ then reduces to the
average unitary operation (\ref{eq:ubar}).
This means that each term of $F''^2_{m_0} -F_{\uni}''^2$
and $P''_{m_0} -P_{\uni}''$ must contain at least one
$\hat{\bar{\epsilon}}^{(2)}$.
Since $\hat{\bar{\epsilon}}^{(2)}$ is of the second order,
we can set
\begin{equation}
\hat{\delta}_{kk'}^{(3)}=\hat{\delta}_{kk'}^{(4)}
=\hat{\epsilon}_k^{(3)}=\hat{\epsilon}_k^{(4)}=0.
\end{equation}
Consequently, in order to calculate
$F''^2_{m_0} -F_{\uni}''^2$
and $P''_{m_0} -P_{\uni}''$ to the order of $g^4$,
it is sufficient to use Eqs.~(\ref{eq:forme}),
(\ref{eq:weakint}), and (\ref{eq:formm}).
Then, we can straightforwardly
derive Eqs.~(\ref{eq:difff}) and (\ref{eq:diffp}).

\section{\label{sec:degf}Degradation of Fidelity}
We here explain the derivation of Eq.~(\ref{eq:fexpress}).
From Eq.~(\ref{eq:defc}),
the weight $|c_\sigma|^2$ in Eq.~(\ref{eq:def1stf}) is given by
\begin{equation}
|c_\sigma|^2=\frac{1}{2^{2s}}
 \frac{(2s)!}{(s+\sigma)!(s-\sigma)!}\,
 \left[\,1+(-1)^{s-\sigma}\,(c_+c_-^\ast +c_+^\ast c_-)\, \right].
\end{equation}
When $s$ is large, the term with factor $(-1)^{s-\sigma}$
can be ignored, since it is canceled by the summation over $\sigma$
due to
\begin{equation}
 \sum_\sigma \frac{1}{2^{2s}}
 \frac{(2s)!}{(s+\sigma)!(s-\sigma)!}\,(-1)^{s-\sigma}\,\sigma^n=0
\end{equation}
for $n=0,1,\ldots,2s-1$.
The weight $|c_\sigma|^2$ is then a binomial distribution
whose mean and variance
are $0$ and $s/2$, respectively.
The central limit theorem
(or equivalently Stirling's formula) states that as $s$ increases,
the binomial distribution becomes close to
a normal distribution with the mean and variance unaltered:
\begin{equation}
|c_\sigma|^2 \simeq \frac{1}{2^{2s}}
 \frac{(2s)!}{(s+\sigma)!(s-\sigma)!} \to
   \frac{1}{\sqrt{\pi s}}\exp\left(-\frac{\sigma^2}{s}\right).
\end{equation}
At the same time,
the summation over $\sigma$ is replaced with the integral over $\sigma$,
\begin{equation}
 \sum_{\sigma=-s}^{s} \to \int_{-\infty}^{\infty} d\sigma,
\end{equation}
since $\sigma$ is now considered as a continuous variable
from $-\infty$ to $\infty$.
Using Eq.~(\ref{eq:approxn}),
the fidelity (\ref{eq:def1stf})
can be written as
\begin{align}
 F'^2 &\sim \frac{1}{\pi s} \int \int d\sigma d\sigma'
 \exp\left[-\frac{1}{s}(\sigma^2+\sigma'^2)\right]  \notag \\
  &\qquad \times \exp\left[-g^2j\sin^2\theta\,(\sigma-\sigma')^2-
  i2gj\cos\theta\,(\sigma-\sigma') \right],
\end{align}
which gives Eq.~(\ref{eq:fexpress}) through the Gaussian integral.


\end{document}